\documentclass[twocolumn,preprintnumbers,amsmath,amssymb,aps, prb,floatfix,groupedaddress]{revtex4}
\usepackage{graphicx} 
\usepackage{dcolumn}  
\usepackage{bm}       
\usepackage{epsfig}
\usepackage{comment}
\usepackage{color}
\usepackage{mathtools}
\usepackage{multirow}
\usepackage{pgfrcs}

\begin{document}
\title{van Hove singularities and spectral smearing in \\ 
high temperature superconducting H$_3$S }

\author{Yundi Quan}
\affiliation{Department of Physics, University of California Davis,
 Davis, California 95616, USA}
\author{Warren E. Pickett}
\email{wepickett@ucdavis.edu}
\affiliation{Department of Physics, University of California Davis,
 Davis, California 95616, USA}

\date{\today}

\begin{abstract}
The superconducting 
phase of hydrogen sulfide at T$_c$=200 K observed by Drozdov and collaborators at 
pressures around 200 GPa is 
simple bcc $Im{\bar 3}m$ H$_3$S, predicted beforehand by Duan {\it et al.}, has
experimental confirmation.  
The various ``extremes'' that are involved -- pressure, implying extreme reduction of 
volume, extremely high H phonon energy scale around 1500K, extremely high temperature 
for a superconductor -- necessitates a close look at new issues raised by these 
characteristics in relation to high T$_c$. We use first principles methods to analyze the H$_3$S 
electronic structure, particularly the van Hove singularities (vHs) and the effect of sulfur.  
Focusing on the two closely spaced vHs
near the Fermi level that give rise to the impressively sharp peak in the density of states,
the implications of strong coupling  Migdal-Eliashberg theory are assessed. 
The electron spectral density smearing due to virtual phonon emission and absorption
needs to be included explicitly to obtain accurate theoretical predictions and
current understanding.
Means for increasing
T$_c$ in H$_3$S-like materials are addressed. 
\end{abstract}
\maketitle
\date{\today}

\section{Introduction}
The recent discovery of superconducting hydrogen sulfide under high pressure by Drozdov and collaborators\cite{droz1,droz2,droz3}, and remarkably  predicted a year earlier by Duan {\it et al},\cite{duan} has reinvigorated the quest for room temperature superconductivity. The predicted structure has been confirmed by x-ray diffraction studies by Shimizu that show that sulfur lies on a bcc sublattice;\cite{shimizu} the protons cannot be seen in x-ray diffraction. The resistivity transitions were also confirmed by Shimizu.  The experimental reports indicate critical temperatures up to T$_c$=203 K in the pressure range of 200 GPa, based on the resistivity transition, the effect of magnetic field on T$_c$, on a H isotope shift of the right sign and roughly the expected magnitude,\cite{droz1} and most recently the Meissner effect has been demonstrated.\cite{droz2} 

\begin{figure}[!ht]
\includegraphics[width=0.8\columnwidth]{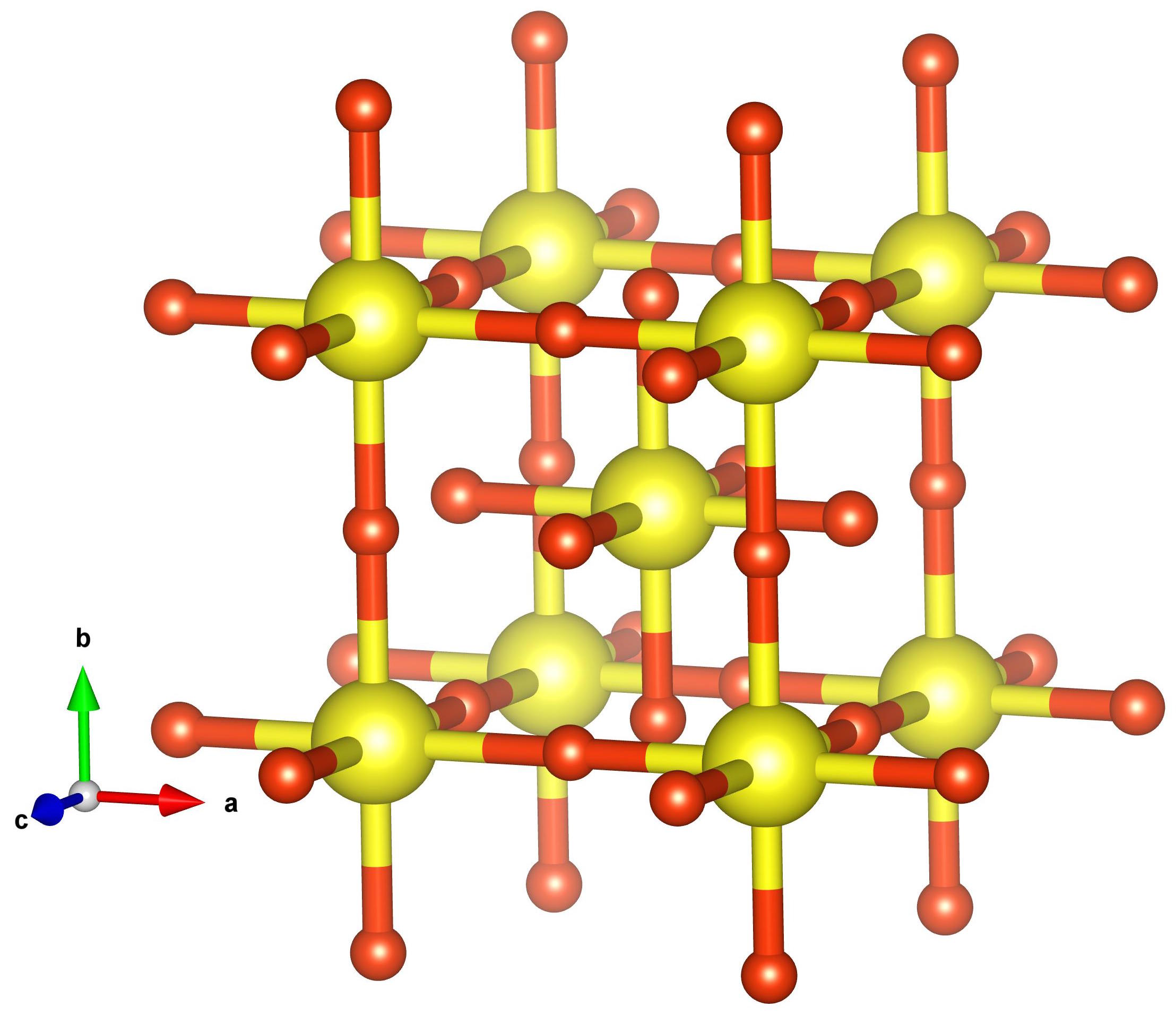}
\caption{Crystal structure of $Im{\bar 3}m$ H$_3$S. Nearest neighbor S-H bonds are shown. At this nearest neighbor level, the structure consists of two interleaved ReO$_3$ sublattices, displaced relative to one another by the body-centering vector (1,1,1)$a$/2.}
\label{structure}
\end{figure}

In a success of predictive theory in this area, the magnitude of T$_c$ in the 200 GPa pressure range was obtained from first principles calculation {\it prior to experiment}\cite{duan} and confirmed by others,\cite{errea,papa,flores} so there can be little doubt that 200 K superconductivity has been achieved in the structurally simple compound H$_3$S, pictured in Fig. \ref{structure}. The finding that H vibrations provide the mechanism seems to confirm the suggestion of Ashcroft that dense hydrogen should superconduct at high temperature,\cite{ashcroft} however evidence is increasing that H-rich materials\cite{ashcroft2} are substantially different and more promising than pure hydrogen until TPa pressures can be reached.
Early quantitative estimates\cite{papa2} of T$_c$ for metallic H were in the 250K range; more
recent values\cite{McMahon} in the range of several TPa lie in the 500-750K range. The phase
diagram of this system is uncertain, however, due to the quantum nature of the proton.

Although comprehensive calculations based on density 
functional theory (DFT) linear response formalism
and Eliashberg theory\cite{SSW} have been reported and seem 
convincing, H$_3$S turns out to be more intricate 
than the initial reports suggest. Using their self-consistent 
harmonic approximation, Errea {\it et al.} find substantial 
corrections due to anharmonicity:\cite{errea} at 200 GPa, 
anharmonicity increases the characteristic
frequency $\omega_{log}$ by 3\%, the electron-phonon interaction
(EPI) strength $\lambda$ is decreased by 30\% and 
the predicted value of T$_c$ falls
22\% from 250 K to 194 K.  Potentially important for further 
understanding is their finding that
anharmonicity shifts coupling strength to H-S bond stretch modes, 
from H-S bond bending (alternatively, H-H bond stretch) modes. 

Other basic questions have yet to be addressed. 
First, why are the electron-phonon matrix elements as large as 
they are? It is true that the main causative 
property behind the high T$_c$ is the (understandably) high phonon 
frequencies that set the energy scale for T$_c$, but substantial 
electron-ion matrix elements are also required. Second, 
Flores-Livas {\it et al.}\cite{flores} have investigated the 
energy dependence of the spectrum around the Fermi level, 
finding that it influences the theoretical 
predictions, which are overly optimistic when energy dependence is
neglected. Both Akashi {\it et al.}\cite{Akashi} and Flores-Livas 
{\it et al.} have solved the gap equations, providing theoretical predictions of
the gap as well as T$_c$ without using the Allen-Dynes equation.
This question posed by intricacies in the density of states (DOS) 
and the role of zero point vibrations has stimulated work by 
Bianconi and Jarlborg.\cite{Bianconi}

More fundamentally there is the question ``why H$_3$S? why sulfur?" Several H-rich 
materials have been studied at high pressure (see references in 
Refs. [\onlinecite{duan,errea,papa,flores}] and Bernstein {\it et al.}\cite{bernstein}), 
and although some are predicted to superconduct up to several tens of kelvins, 
H$_3$S is a singular standout. Li {\it et al.}, for example, studied the H$_2$S stoichiometry for
stable compounds up to similar high pressures,\cite{Li} finding a maximum T$_c$ of ``only'' 80 K.
There is little understanding so far of the microscopic cause of very high T$_c$, beyond 
the obvious expectation of higher phonon frequencies at high pressure; 
the origin of the large matrix elements remain obscure. Papaconstantopoulos and 
collaborators\cite{papa} calculated the pressure dependence of matrix elements, finding increasing H 
scattering with increasing pressure. More basically one can ask, is there something special about sulfur, 
and the underlying electronic structure, that provides the platform for such high T$_c$? 

It is the last of these questions we address initially in this paper. An obvious feature for study is the strikingly sharp peak in the density of states $N(E)$ due to two van Hove singularities (vHs) separated by 300 meV very near the Fermi level $E_F$. There is a large literature on the connection between peaks in $N(E)$ and high T$_c$ in the A15 class of materials\cite{Labbe} and later in the high temperature superconducting cuprates,\cite{BobM,Abrikosov} but their importance for H$_3$S is unclear. van Hove singularities near the Fermi level can enhance N($E_F$) and thus the EPC strength $\lambda$ due to increased number of available states to participate, but there are additional questions to address.

The paper is organized as follows. Methods are described in a brief Sec. II. In Sec. III the general electronic structure and the charge density near $E_F$ are presented and discussed. Based on a Wannier function representation of the bands, a minimal tight binding model is presented, with the intention of identifying the important features of the bonding and especially the deviation from free electron like density of states over much of the valence band. The two van-Hove singularities are identified, quantified, and analyzed, and the relation between them is identified. In Section IV we address the peak in N(E) in the light of strong electron-phonon coupling, high frequencies, and thermal smearing. Sec. V presents scenarios for further increase in coupling strength $\lambda$, and raising of T$_c$ toward room temperature, in this and similar systems. A short Summary is provided in Sec. VI.

\section{Methods}
Density functional calculations have been carried out using both the linearized augmented plane wave (LAPW) method based WIEN2k code\cite{Wien2k} and the linear combination of atomic orbitals based FPLO code.\cite{fplo} The PBE implementation\cite{PBE}  of the generalized gradient approximation (GGA) is used as the exchange correlation functional. The crystal structure of
H$_3$S is $Im\bar{3}m$ with a lattice constant of 5.6 a.u. corresponding\cite{papa}
 to a pressure of 210 GPa.

\begin{figure}[!ht]
\includegraphics[width=0.8\columnwidth]{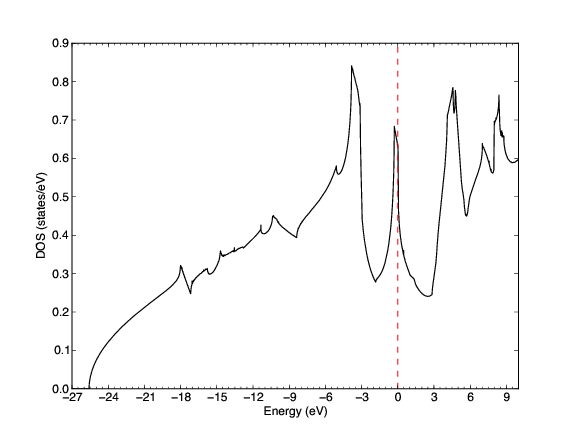}
\vskip 4mm
\includegraphics[width=0.7\columnwidth]{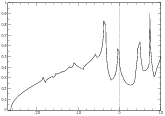}
\caption{Full density of states for WIEN2k (top panel) and FPLO (bottom panel) calculations.
 The bandwidth is extremely large due to the broadening caused by short
 interatomic distances. The region of the van Hove singularities differ in
detail.  $N(E_f)$ from FPLO is 0.44 states/eV, while from WIEN2k it is 0.63 states/eV.}
\label{fullDOS}
\end{figure}

To study the vHs points near the Fermi level, a very fine k-mesh containing 8094 (8112) points for WIEN2k (FPLO) in the irreducible Brillouin zone is used. Sphere radii R for H and S are 0.97 $a_0$ and 1.81 $a_0$ respectively, with basis set cutoff determined by R$_H$K$_{max}$ = 6. The results we discuss are insensitive to these choices. The tight-binding parameters we present were obtained as a two-center Slater-Koster simplification of a more extensive representation in terms of symmetry-adapted Wannier functions as implemented in FPLO. 

\section{Electronic structure and bonding}
\subsection{Electronic structure at 210 GPa}
The calculated lattice constant\cite{papa} at 210 GPa, $a$=5.6 a.u. corresponding
to a volume 58\% of the zero pressure volume in the same structure, is used in
all calculations. The DOS $N(E)$ on a broad scale from the 
two all-electron, full potential codes are presented in Fig. 2. Because fine 
structure is of interest here, we compare a few results from two all-electron, 
full potential codes. The occupied bandwidth is 26 eV. Over the lowest 20 eV 
of this range, the DOS has a remarkably free-electron-like $\sqrt{E}$ shape, 
without significant structure. Over the lower end of this region, the DOS is 
dominated by S $3s$ character, above which H $1s$ and S $3p$ character enter and 
mix. Then, at -4 eV and +5 eV two substantial and rather narrow peaks emerge, 
indicative of very strong hybridization, perhaps bonding and antibonding signatures. 
Double valleys lie at -2 eV and +2 eV, between which a very sharp peak, related 
to two van Hove singularities (vHs) 0.25-0.30 eV apart in energy, juts upward. 
The Fermi energy $E_F$ (set to zero throughout) lies very near the upper vHs. 

More details of the vHs region from WIEN2k and FPLO are left to Appendix A. 
Due to relatively small differences but a very sharp peak, the values of 
N(E$_F$) differ substantially:
0.63/eV-f.u. from Wien, 0.44/eV-f.u. from FPLO.
Pseudopotential results will likely give similarly differing values.  
These differences  
arise because the input parameters (orbitals, sphere radii, pseudopotentials,
cutoffs, etc.) may not be optimized for application at such reduced volumes. 
We demonstrate in Sec. IV that for physical superconducting properties, thermal 
and dynamical broadening makes details of
N(E) fine structure relatively unimportant. This unimportance {\it does 
not however apply} for the underlying theory, where it has serious consequences
(discussed in Sec. V) partly because so much is formulated and
evaluated in terms of the specific value of N(E$_F$)
but also because the energy dependence has significant impact.

Returning to the DOS, such strong structure in N(E) reflects strong mixing between orbitals lying in this energy range, which are the H $1s$ and S $3p$ valence orbitals. The orbital projected DOS (PDOS) presented by Papaconstantopoulos {\it et al.}\cite{papa} shows that S $3s$ participation is becoming small around E$_F$. Their PDOS helps to understand the strong DOS structure. The peak at -4 eV is largely S $3p$ character with some H $1s$ contribution. The peak at +5 eV has, surprisingly, a large contribution from Bloch orbitals with $d$ symmetry around the S site, with some participation of all of the orbitals besides S $s$. The peak at E$_F$ -- the important one bounded by two vHs -- is a strong mixture of H $1s$ with S $3p$, whose corresponding tight binding hopping parameters will have a correspondingly large hopping amplitude.

\subsection{Minimal tight-binding model}
In this section, a minimal tight-binding model for $H_3S$ will be constructed using Slater-Koster two center parameters. Local basis orbitals are the atomic orbitals of sulfur $S$, $P_x$, $P_y$, $P_z$ and the three hydrogen $1s$ orbitals ($s$); this notation is used below. The procedure is to first calculate the symmetry projected Wannier functions (WFs), based upon the seven basis functions. The resulting WFs at reasonably large density isocontour (shown in Appendix B) reflect hybridizing atomic orbitals. In the generation of the WFs, a set of three center hopping integrals is generated, with the WFs as the basis orbitals. From the WF three center integrals, simpler two center integrals can be obtained. Some members of this latter set may be overdetermined, in which case a best choice (average) must be made.

The on-site energies and largest Slater-Koster parameters are listed in Table I. Relative to E$_F$=0, the on-site energies (compared to those reported by Bernstein {\it et al.}\cite{bernstein}, in parentheses) are: $\varepsilon_S$ =-8.0 (-8.6) eV; $\varepsilon_s$ =-5.5 (-5.0) eV; $\varepsilon_P$ =~0.0 (-1.3) eV. The procedures used by Bernstein {\it et al.} are not exactly the same as ours, with the difference indicating the level of confidence one should assign to these energies considering the non-uniqueness of tight binding representations. It is eye-catching that our sulfur P on-site energy is indistinguishable from E$_F$. 

\begin{figure}[!ht]
    \includegraphics[width=\columnwidth]{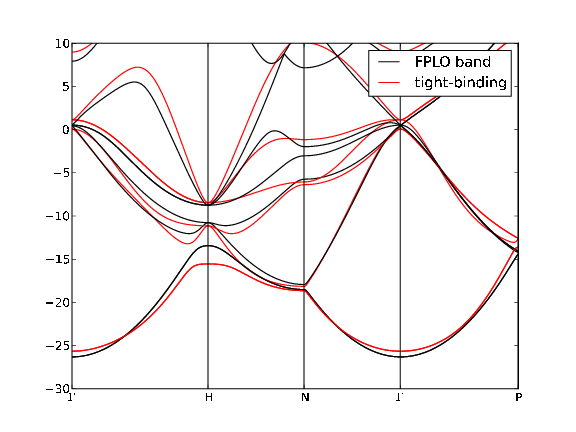}
    \caption{Bands from the minimal two-center tight-binding model described in the text (red lines), compared 
  with the full band structure (black lines). The agreement is reasonable but approximate over this 30 eV region.}
    \label{TBbands}
\end{figure}

 \begin{table}[!ht]
 \caption{Slater-Koster two-center parameters (in eV).The subscripts indicate 
the index of the neighbor: 1$\equiv$1st neighbor, 2$\equiv$2nd neighbor, etc. 
For H there are two 4th neighbors one lattice constant apart: one through the S 
atom, denoted ``4,'' one through another H atom, denoted ``4$^{\prime}$.'' 
$(\widetilde{PP})_1$ indicates $\frac{1}{4}(PP\sigma)_1 + \frac{3}{4}(PP\pi)_1$, 
which only occurs in this combination.}
 \begin{tabular}{cc|cc}
 \hline
$\epsilon_S$        & -7.98     &  $(sP\sigma)_1$ & -5.42  \\
 $\epsilon_s $        & -5.46 & $(PP\sigma)_2$ & -1.83   \\
 $\epsilon_P $        & -0.03 & $(SP\sigma)_2$ & 1.29\\ 
 \cline{1-2}
  $(sS\sigma)_1$         & -4.37 & $(SS\sigma)_2$ &  0.94    \\
$(ss\sigma)_1$           & -2.80 & $(sP\sigma)_2$ & -0.93\\ 
$(ss\sigma)_4$           & -1.14 & $(\widetilde{PP})_1$ & 0.60 \\ 
  $(ss\sigma)_4^\prime$  &  0.55 & $(SS\sigma)_1$ & 0.30     \\
 \hline
 \end{tabular}
 \label{slater-koster}
 \end{table}
 The largest hopping parameters are nearest neighbor (n.n.)  $sP\sigma$ (-5.4 eV) and $sS\sigma$ (-4.4 eV), 
leading to the possibility that H-S n.n. hopping is dominant in creating the DOS structure. We have 
calculated the DOS for a single sublattice, ReO$_3$ structure H$_3$S, and it is nothing like the full DOS.  
The n.n. H-H hopping $ss\sigma$, at -2.8 eV, couples the sublattices strongly, and S-S second neighbor hoppings,
one lattice constant apart,  are 1.3-1.8 eV in magnitude. Interesting are the 4th neighbor H-H hoppings, 
between atoms one lattice constant apart. The hopping through the S atom $(ss\sigma)_4$ is -1.1 eV, 
while through an intervening H atom $(ss\sigma)'_4$ is half that size with opposite sign.  The bands from 
this `minimal' tight binding model are compared with the DFT bands in Fig. \ref{TBbands}, where it can be seen 
that it captures the general behavior (but not the vHs) of the full band structure over a 30 eV range. Many more hopping 
parameters are necessary to reproduce the band structure accurately.
This result suggests that a tight binding representation is not natural for H$_3$S.

\subsection{Role of Sulfur}
 What then is the role of S? We have taken another view of this issue, by comparing H$_3$S with $H_3$H, 
{\it i.e.} the S atom replaced by another H atom. This is simply simple cubic hydrogen with a lattice constant 
$a$=2.8$a_{\circ}$, but we calculate it as H$_3$H for comparison and uniformity. The DOS and band structure 
are presented in Appendix C. The DOS bears little resemblance to that of H$_3$S. The occupied bandwidth is 
15 eV, and the lower 10 eV of this is free electron like.  A vHs is encountered at -2 eV followed by a 
remarkably linear N(E) over almost 10 eV. Then a second vHs signals the minimum of another free-electron-like 
high conduction band. This confirms that it is strong H-S $sS\sigma$ and
 $sP\sigma$ bonding that produces the strong structure in N(E) shown in Fig. 2. Thus sulfur is crucial
in producing the DOS peak at E$_F$ in H$_3$S. We note that the first vHs does not appear below E$_F$ in
either\cite{papa2} fcc or bcc H, which are closer packed and more stable phases of elemental H.
 
The peak in N(E)  reveals that the band filling in H$_3$S happens to be almost perfect (E$_F$ lying very 
nearly at the sharp peak). Retaining the same band filling suggests substituting Se or Te for S. Flores-Livas {\it et al.}\cite{flores} have done parallel calculations for H$_3$S and H$_3$Se. The H$_3$Se frequency $\omega_{log}$ is 10\% {\it higher} but the calculated value of $\lambda$ is lower by 40\%, with the resulting T$_c$ being lower by 27\%.
The changes of $\omega_{log}$ and $\lambda$ indicate that the product $\eta = N(E_F)<I^2>$ is lower by 20\% for the Se compound. Here $<I^2>$ is the Fermi surface average of the square of the electron-H ion scattering matrix element. With H so dominant in the EPI and H modes separated from S (or Se) modes, the picture is dominated by
\begin{eqnarray}
\lambda_H = \frac{N(E_F)<I_H^2>}{M_H \omega_H^2},
\end{eqnarray}
where the matrix element refers to scattering from the displaced H potential and $\omega_H$
is a characteristic frequency from H modes.

The other isovalent ``chalcogenide'' is oxygen, which is quite different from S chemically with H. 
We have found that the DOS of H$_3$O in the H$_3$S structure differs substantially from that of H$_3$S. It may be relevant that
H$_2$O does not metalize until {\it much higher} pressures than are being considered here. Heil and
Boeri\cite{Heil} have considered bonding, EPI, and T$_c$ where sulfur is alloyed with other group VI
atoms. With alloying treated in the virtual crystal approximation (averaging pseudopotentials), 
they have suggested that a more electronegative ion will help. This leaves only oxygen in that column,
 and they calculated that a strong increase in matrix elements compensates a considerable decrease
in N(E$_F$), so that  $\lambda$ might increase somewhat. Ge {\it et al.} have also suggested partial
replacement\cite{Ge} of S, with P being the most encouraging, due to the increase in $N(E_F)$; however,
spectral density smearing will decimate this difference. 

While S changes the electronic system very substantially from that of H$_3$H, it may not be
so special. A variety of calculations have predicted 
(see Durajski {\it et al.}\cite{Durajski} for references)
high values of T$_c$ (in parentheses) for H-rich solids: SiH$_4$(H$_2$)$_2$ (107 K at 250 GPa),
B$_2$H$_6$H (147 K at 360 GPa), Si$_2$H$_6$ (174 K at 275 GPa), CaH$_6$ (240 K at 150 GPa). Whether
any general principles can be extracted from these results remains to be determined.

\subsection{Charge density within 1eV of Fermi level}
\begin{figure}[!ht]
\centering
\includegraphics[width=0.8\columnwidth]{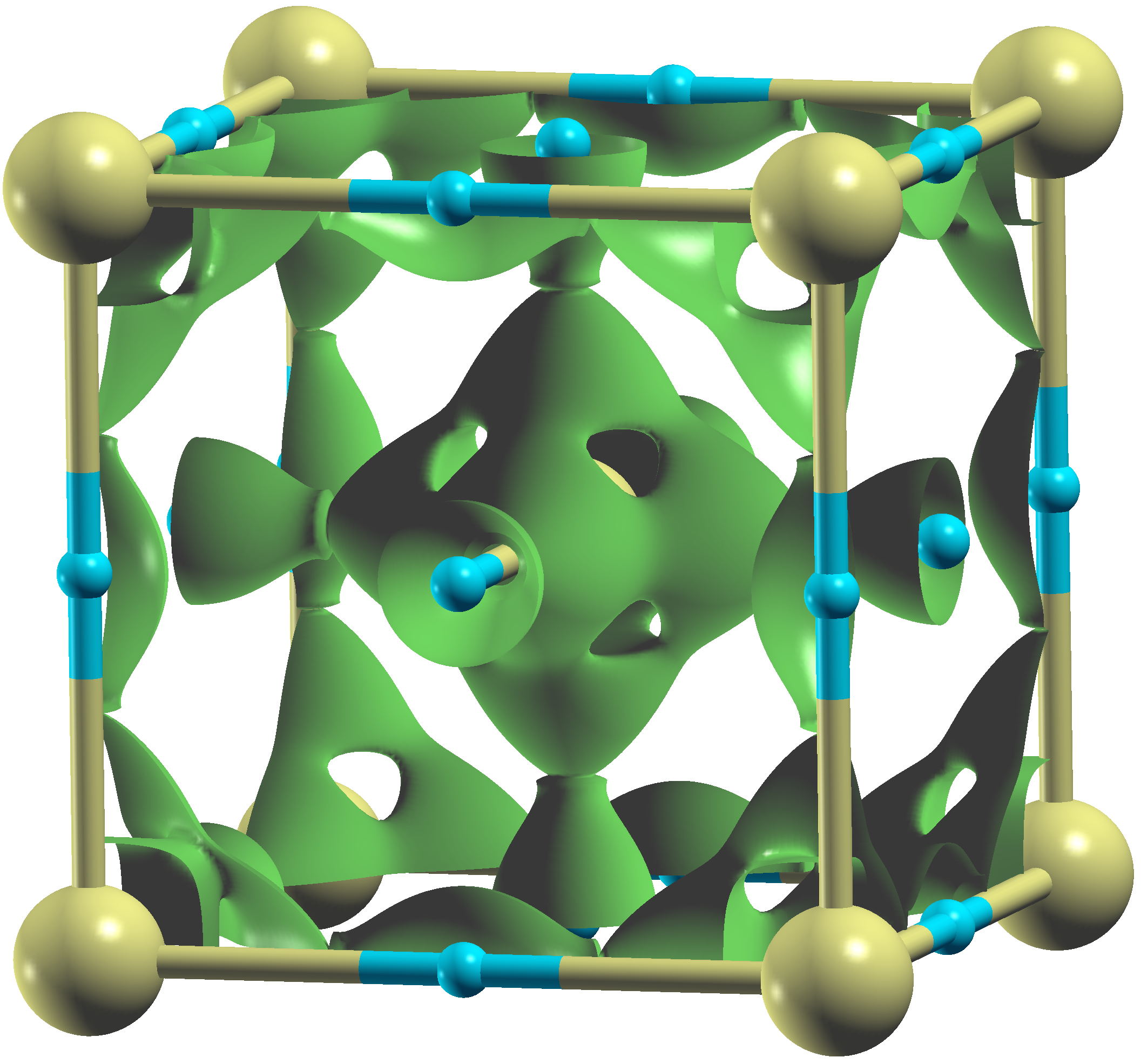}
\caption{Isosurface of the charge density obtained from states in the energy range E$_F$$\pm$1 eV. Sulfur atoms (yellow spheres) lie at the corners and at the body center, H atom are denoted by small blue spheres.}
\label{fig:chgden}
\end{figure}

The charge density from states within 1 eV of E$_F$ is shown in Fig. 
\ref{fig:chgden}; it is this density whose coupling to H vibrations
gives strong coupling and the very high value of T$_c$. Results from 
smaller energy slices are no different, indicating that the states in this 
range have the same character. Two of the bands become flat in a region away from symmetry lines, and the resulting two vHs give rise to the sharp and narrow peak. 
The density around S is strongly distorted from spherical symmetry, having 
substantial maxima in the direction of neighboring H atoms. The H density 
is strongly elongated toward the two neighboring S atoms, more strongly than might have been guessed. These shapes 
reflect strong covalent H $1s$ - S $3p_{\sigma}$ interaction, although there 
remains a density minimum in the bond center rather than a bond 
charge maximum. This character is typical of strong directional bonding 
in metallic compounds. A more nuanced indication of hybridization is available from the isosurface plots of the symmetry-projected Wannier functions presented in the Appendix B.

\subsection{van Hove singularities of H$_3$S}
N(E)  expanded in the energy range -1.0 eV to 0.2 eV is plotted in Fig. 7 in Appendix B,
which reveals the two vHs at $\varepsilon_{lo}$ and $\varepsilon_{hi}$ from both
the Wien2k and FPLO codes. 
Fitting N(E) to
the following piecewise expression\cite{Mecholsky} for 3D vHs near two singularities
\begin{equation}
N(E) =  
\begin{dcases}
a_1 \sqrt{|b_1-\epsilon|}+c_1\epsilon+d_1 &   \epsilon < \epsilon_{lo}
\\
a_2 \epsilon+b_2         & \epsilon_{lo} <  \epsilon < \epsilon_{hi}
\\
a_3\sqrt{|\epsilon+b_3|}+c_3\epsilon+d_3 & \epsilon_{hi} <  \epsilon 
\label{eqn1}
\end{dcases}
\end{equation}
we obtain an excellent fit as can be seen in Appendix B, Fig. 7, 
demonstrating that  contributions from other bands 
are smooth and slowly varying on this energy scale.

\begin{figure}[!ht]
\includegraphics[width=\columnwidth]{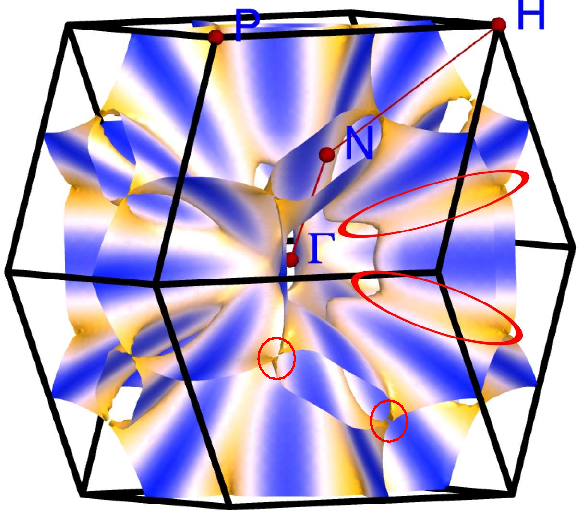}
\includegraphics[width=\columnwidth]{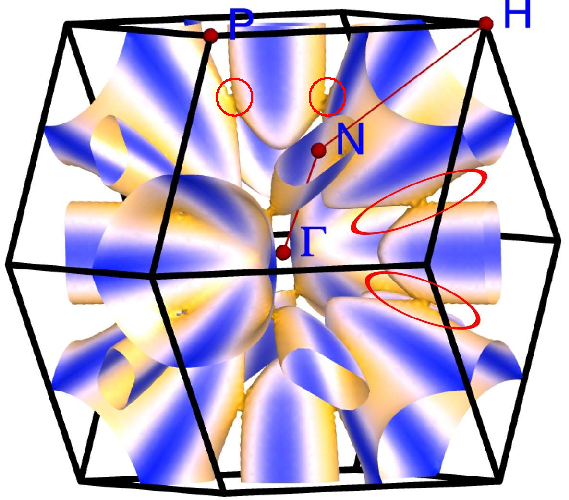}
\caption{Top: iso-energy contour from FPLO for the vHs at E=-0.43eV. Bottom: similarly for the vHs at E=-0.11 eV. The red circles pinpoint two of the symmetry related vHs in each case. The red ellipse outlines the region joined by the two van Hove singularities; two energy surfaces are ``unzipped' as the energy increases between the two vHs. Color denotes the velocity, which ranges from zero (darkest tan, at the vHs) to 2.5$\times 10^8$ cm/s (deep blue).}
\label{fig:vHs}
\end{figure}

Isosurfaces at the two vHs energies are presented in Fig. \ref{fig:vHs}. 
The vHs points lie at either end of a line where low velocity regions of 
two sheets of Fermi surface break apart at the zone boundary, and then 
``unzip'' until they separate into  disjoint sheets. The vHs (from 
the FPLO bands) at -0.43 eV occurs at (-0.42,0.21,0)$\pi/a$ and symmetric points. 
In the local principal axis coordinate system the effective masses are 
-0.15$m_e$, 1.36$m_e$, 0.14$m_e$, giving a thermal (or DOS) mass 
$m_{th} \equiv |m_1 m_2 m_3|^{1/3}$ = 0.31$m_e$. For the one at -0.11 eV, 
the masses are -0.83$m_e$, -0.16$m_e$, 0.56$m_e$, and $m_{th}$=0.42$m_e$.
We return to the importance of vHs effective masses in Sec. VI.

\section{Impact of electron-phonon coupling }
Thermal broadening as described by the Fermi-Dirac distribution function 
has well understood consequences, and is a factor in determining (hence 
limiting) T$_c$, though it is usually not thermal broadening of the DOS, 
which is normally constant over an energy range of several $k_BT_c$. 
The purely thermal aspect is formalized in the gap to thermal broadening 
scale ratio $2\Delta/\pi k_B T_c \approx$ 1, and is central to, but standard in, 
Eliashberg theory. Specifically, it is thermally excited electron-hole 
excitations that overcome pairing and restores the normal state above T$_c$. 

However, the sharp structure in N(E) on the scale of relevant phonon energies requires an extension of conventional implementations of Eliashberg theory,\cite{SSW} where a constant N(E) on the phonon energy scale is assumed, so all scattering processes can be considered as confined to the Fermi surface E=E$_F$. This issue was confronted long ago,\cite{WEP0,WEP1} because of the sharp structure in N(E) in several of the 
then-high-T$_c$ A15 structure compounds, viz. Nb$_3$Sn, V$_3$Si, Nb$_3$Ge, 
with T$_c$$\sim$20 K, and has been followed up in related 
applications.\cite{Mitrovic,Radtke} 

In H$_3$S around 200 GPa, the representative frequency is $\Omega$ $\sim$ 1300 K = 112 meV.
We use this value below, based on the DFT-based calculations of (harmonic) $\omega_{log}$
of 1125 K [\onlinecite{Heil}],
1264 K [\onlinecite{errea,flores}],
1335 K [\onlinecite{duan,Akashi}], and 1450 K [\onlinecite{Ge}].
The logarithmic, first, and second moments differ only at the 2\% level, 
whereas reported values differ 
by as much as 20\%. Using a somewhat different value of $\Omega$ below 
would not change our conclusions.

From visual examination, the sheets of the constant energy surfaces at 
E$_F$$\pm$$\Omega$ do not differ much from the Fermi surfaces at E$_F$ 
(see Fig. 5), the differences occurring in small pockets around $\Gamma$ 
(not visible) and along a line connecting the two vHs, 
irrespective of the electronic structure 
code that is used. Inter-vHs scattering could be interesting: although it 
involves a small amount of phase (q) space, it incorporates a disproportionate 
fraction of states with low to vanishing velocity. Possible complication from 
inter-vHs scattering and non-adiabatic processes lie beyond the scope of our discussion.

\subsection{Formalism}

It has been known since the work of Engelsberg and Schrieffer\cite{epi1} and 
Shimojima and Ichimura\cite{epi2} that, for a characteristic phonon frequency $\Omega$
in an interacting electron-phonon system, electron spectral density is spread from
its noninteracting $\delta$-function spike at $E_k$ up to a few $\Omega$.
The spectral density arises from the electronic self-energy that is treated for
superconducting systems by Eliashberg theory.
When N(E) hardly varies over a scale of  a few $\Omega$ it is rare to notice the effects of
such broadening except possibly in direct measurements where phonon sidebands may be observed in
photoemission spectra.
For situations as in H$_3$S where Bloch state character is slowly varying in energy 
but N(E) varies rapidly, the normally simple electron-phonon formalism becomes
challenging. Drozhov studied the effects of EPI in the vicinity of a vHs where Migdal's
theorem is violated, and found severe renormalizations.\cite{drozhov} If these are
confined to a very small phase space, however, the effects on most properties may be minor.

For the case of rapidly varying N(E) but neglecting violations of Migdal's theorem,
the generalization of Eliashberg theory has been formulated and applied to the A15 
compounds.\cite{WEP1,WEP2} One feature that is distinctive in H$_3$S compared to most
other EPI superconductors is that T$_c$ 
is an order of magnitude higher, because the frequencies are comparably higher, and 
simple thermal broadening is correspondingly larger and requires attention. The 
second factor in common, and the important one, is that strong EPI causes an 
effective smearing of the electronic spectral density due to exchange of virtual 
phonons appearing in Migdal-Eliashberg theory -- excitations described by the electron Green's
function are part electron, part phonon. 
This broadening is given by the imaginary part of the interacting electronic Green's function
\begin{eqnarray}
G_k^{-1}(\omega)~=~\omega-[E_k-\mu(T)]    -M_k(\omega;T)-i\Gamma_k(\omega;T).
\end{eqnarray}
Here $E_k$ is the DFT band energy, $\mu(T)$ is the chemical potential, and $M$ and $\Gamma$ are the real and imaginary parts of the phonon-induced self-energy.

The spectral density $A(\omega)$ is the interacting analog of the band DOS N(E):
\begin{eqnarray}
A(\omega)&=&\sum_k A_k(\omega) = \frac{1}{\pi}\sum_k |Im G_k(\omega)| \\ \nonumber 
         &=&\frac{1}{\pi}\sum_k \frac{\Gamma_k(\omega)}
            {[\omega-(E_k-\mu)-M_k(\omega)]^2
            + \Gamma_k(\omega)^2}   \\ \nonumber
        &\rightarrow&  \int d\xi 
    \frac{\Gamma/\pi}{(\omega-\xi)^2 + \Gamma^2}N(\xi).
\end{eqnarray}
In the last expression the Brillouin sum has been converted into an energy integral by inserting $\int dE \delta(E-E_k)$=1, assuming that only $E_k$ (and not wavefunction character, hence not $M$ or $\Gamma$) depends on $k$ near E$_F$, and $\xi_k = \varepsilon_k - \mu + M(k,\xi_k)$ is the quasiparticle energy. This simplification is usually fine for electron-phonon coupling in a standard Fermi liquid, 
as wide-band  H$_3$S appears to be. 

There is strong rearrangement of spectral density even before this smearing effect of
electron damping $\Gamma$. For temperatures and frequencies $\omega$ up to the order of the
characteristic phonon energy $\Omega$ or more, the behavior of the real part $M_k$ is
linear $dM_k/d\omega = -\lambda_k$, where $\lambda_k$ is the EPI strength at $k$
whose average over the Fermi surface is $\lambda$. The equation for $\xi_k$ in the
previous paragraph then gives for the quasiparticle energy 
\begin{eqnarray}
\xi_k = \frac{E_k -\mu}{1+\lambda_k}.
\end{eqnarray}
This equation expresses the phonon-induced mass enhancement, and $(1+\lambda_k)^{-1}$ is
the quasiparticle strength, i.e. the fraction of the electron's 
$\delta$-function spectral density at
$\xi_k$ and whose average in H$_3$S is 1/3 ($\lambda\approx 2$). Two-thirds of the spectral
weight is spread from $\xi_k$ by up to a few times $\Omega$. This is a serious 
redistribution of weight that we cannot treat in any detail without explicit solution
for the self-energy on the real axis.

Notwithstanding the complications, in an interacting system the thermal distribution 
function containing all complexities can be handled formally to provide insight
into this ``varying N(E)'' kind of system. The interacting thermal distribution
(state occupation) function $f(E_k)$ is defined as the thermal expectation of 
the number operator $n_k$
\begin{eqnarray}
f(E_k)&=&T\sum_{-\infty}^{\infty} G_k(i\omega_n) e^{i\omega_n\eta} = 
  \int_{-\infty}^{\infty} d\omega f_{\circ}(\omega)A(E_k,\omega),  \nonumber
\end{eqnarray}
where the Matsubara sum with positive infinitesimal $\eta$ has been 
converted into an integral in the 
last expression and $f_{\circ}(E)$ is the (non-interacting) Fermi-Dirac  distribution. 
The interacting distribution function can be expressed as the 
non-interacting one broadened by\cite{WEP2} 
$\Gamma_k(\omega)$ as N(E) is broadened in Eq. (4). 

Several thermal properties can be 
formulated\cite{WEP2} in terms of the 
interacting  (broadened and in principle mass renormalized) density of states ${\cal N}(E)$. 
Returning to single particle language, the spectral density at $E_F$ is approximately
\begin{eqnarray}
{\cal N}(E_F)=\int dE \frac{\Gamma/\pi}{(E-E_F)^2 + \Gamma^2} N(E).
\end{eqnarray}
For energies a few $\Omega$ around $E_F$ the extension $E_F\rightarrow E$
to give ${\cal N}(E)$ will be reasonable.
Then, returning to the distribution function, the total electron number can be 
written\cite{WEP2} in two ways
\begin{eqnarray}
N_{el} = \int dE f(E)N(E)=\int d\omega f_{\circ}(\omega){\cal N}(\omega),
\end{eqnarray}
illustrating that interaction effects can be exchanged between the distribution 
function and, in this instance, the interacting and non-interacting density 
of states.
Thus in a region around $E_F$ the spectral density is the band density of states 
broadened by a Lorentzian of halfwidth $\Gamma$. 

\subsection{Thermal and phonon smearing in H$_3$S}
We now estimate the impact of this spectral density smearing for H$_3$S using the
Wien2k result for N(E).
The mass renormalization effects (from the real part of the self-energy)  
Migdal theory are outlined in Appendix D but will be disregarded here, leaving
our estimate as an {\it underestimate} of the effect of smearing. Investigation of the 
(Migdal) self-energy equations ({\it i.e.} in the normal state) gives the quasiparticle inverse 
decay rate via phonons over most of the relevant energy range,\cite{epi1,WEP0,WEP1} for an 
Einstein model, as
\begin{equation}
\Gamma \approx \pi \lambda \Omega \large[n_B(\Omega)+
    \frac{1}{2}\large]
\end{equation}
For H$_3$S the characteristic frequency $\Omega$ is in the ballpark of $\Omega\approx$
1300 K (see above). Since we are interested only in relative values of quantities
affected by smearing,
we will not distinguish
$\omega_{log}$ from $\omega_2$, etc. 

With the choice of $\mu^*$=0.15, a value of
$\lambda$=2.17 is necessary to give the observed T$_c$=200K, which also is in the
range that has been quoted as resulting from DFT calculations.
At 200 K, the Bose-Einstein thermal distribution 
$n_B$ gives a negligible fraction of phonons excited: $n_B(\Omega) \sim 10^{-4}$. Thus
\begin{equation}
\Gamma = \frac{\pi}{2}\lambda\Omega = 5\times 10^3~K = 0.38~{\rm eV},
\end{equation}
the halfwidth is proportional to the product $\lambda \Omega$.
This smearing in Eq. (6) arises from the virtual excitation of phonons that provides the 
coupling, even at low T$_c$ where phonons are not excited. We note that zero-point vibrations 
do not scatter electrons; sharp Fermi surfaces survive strong electron-phonon 
coupling, with de Haas -- van Alphen oscillations remaining visible, and
the resistivity $\rho(T)\rightarrow 0$ at $T\rightarrow 0$. 

\begin{figure}[!ht]
\centering
\includegraphics[width=\columnwidth]{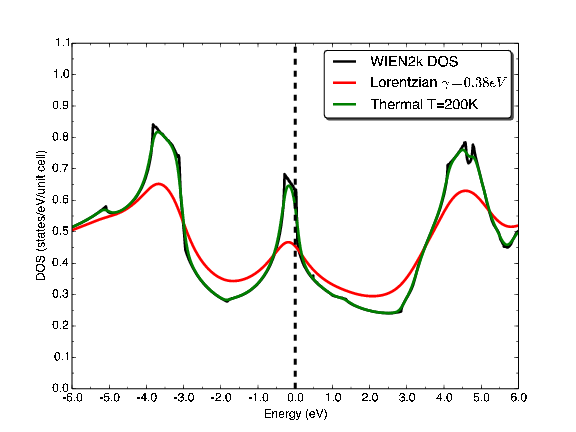}
\caption{H$_3$S density of states without broadening (black line, Wien2k DOS), 
with thermal broadening at 
200 K (green line, which is hard to distinguish from the unbroadened one), and the 
virtual phonon broadened effective DOS {\cal N}($E$)
using a Lorentzian halfwidth of 0.38 eV (red line). 
Note the large drop in the value at the Fermi energy 
(dashed vertical line).}
\label{fig:lorentzian}
\end{figure}

Figure~\ref{fig:lorentzian} shows three curves: the calculated (Wien2k) static lattice 
N(E) from Fig. 2, the thermally broadened version at 200 K, and the virtual phonon 
broadened DOS with halfwidth $\Gamma$=0.38 eV
from above. The thermal broadening function is the derivative 
$-df_{\circ}(E-E')/dE$ with half width of about $\pi$k$_B$T = 625 K = 55 meV. 
This thermal broadening at 200 K is minor on the scale of interest, reducing 
the effective value of N(E$_F$) slightly. The phonon broadening however is severe, 
with the peak value of 0.70/eV-f.u. for N(E) dropping by 37\%.  The unbroadened 
value N($E_F$) of 0.64 states/eV-f.u. is lowered by nearly 1/3, to 0.45 states/eV-f.u. 
The effect of the shift of the chemical potential is secondary when broadening 
is so large.

\subsection{Implications for the theory of H$_3$S}
The Eliashberg equations including the energy dependence\cite{WEP0,WEP1} of N(E) 
indicate that it is this 1/3 reduced value of N(E$_F$) that should be used with 
the standard implementation to get a good estimate of $\lambda$, the 
superconducting gap, and T$_c$. The impact of experiment/theory agreement for 
H$_3$S is substantial and negative: the spectral function $\alpha^2F(\omega)$, 
proportional to $N(E_F)$, is reduced by 1/3 by EPI. The naive value of 
$\lambda\approx$ 2.17 becomes, after reducing by 1/3, $\lambda$=1.45. The phonon 
frequency moments, which involve $\alpha^2F/\lambda$, seem at this stage
to remain unchanged; we return to this point below.

We have evaluated the magnitude of this phonon broadening correction on T$_c$ using the 
Allen-Dynes equation,\cite{alldyn} taking the representative values for H$_3$S of 
the phonon moments to be $\Omega$ = 1300 K and $\mu^*$=0.15. For $\lambda$=2.17, T$_c$ = 200 K;
for the 1/3 reduced value $\lambda$=1.45, T$_c$=130 K. The agreement between theory
and experiment is strongly degraded. It is worthy of note that  H$_3$S happens to be 
in a nearly linear regime of T$_c(\lambda$), where a reduction of $\lambda$ by 
1/3 results in a decrease of T$_c$ by nearly 1/3.  We note that the corresponding 
strong coupling factor $f_1$ in the Allen-Dynes parametrization is 1.13 (1.07); 
$f_1$ is the crucial improvement of the Allen-Dynes equation over the McMillan 
equation and seems sometimes for hydrogen sulfides to have been neglected.  
 
What this comparison implies is that theory-experiment agreement for T$_c$ is
not as good as has seemed, since taking phonon smearing into account, theory
would only be predicting of the order of 2/3 of the ``constant $N(E_F)$'' value. It should be
noted that this 1/3 reduction factor depends on the accurate calculation of
$N(E)$, for which we have used the Wien2k spectrum. With the FPLO result for
the DOS, $N(E_F)$ is lower and thus the effect of smearing will be smaller.

A few papers have reported calculations of the band structure and of N(E), and 
some studies have reported $\lambda$, but little attention has been given to the 
value of N(E$_F$), which is sensitive to method and computational procedures. 
At 200 GPa ($a$=5.6 a.u.) Papaconstantopoulos {\it et al.} quote 0.51/eV-f.u.; 
from the figure of Duan {\it et al.} we estimate the small value of 0.2/eV-f.u. 
(certainly their presented peak is much weaker): Bianconi and Jarlborg report 
0.50/eV-f.u. (their table numbers must be per cubic cell). Our Wien2k and 
FPLO values are 0.64 (0.42)/eV-f.u.
respectively, indicating that even all-electron full potential methods can differ.

The point is that in conventional Eliashberg theory -- constant N(E) on the 
phonon scale -- $\lambda$ is proportional to N(E$_F$), and the values that have 
been used are sensitive to methods and cutoffs (depending on method, see above), 
but more seriously they are obtained from unbroadened N(E). Because of this, 
the reported values of $\lambda$ and hence T$_c$ are quantitatively uncertain, 
assuming they are converged BZ integrals. And  on this point, 
McMahon and Ceperley\cite{McMahon} and Akashi {\it et al.}\cite{Akashi} 
have discussed the various challenges in reaching convergence, before even
confronting the energy variation question. The
(unsmeared) prediction of T$_c$$\approx$ 200 K indicates that improved theory, 
by taking into account phonon broadening, would give a substantially reduced
critical temperature.

Flores-Livas {\it et al.} have recognized the issue of the variation of N(E) 
on the scale of the phonon frequencies. In their implementation of density 
functional theory for superconductors (DFTSC), this variation is accounted for. 
They reported, for their calculation of N(E) (details were  not reported), 
taking into account the variation resulted in a 16\% decrease in T$_c$, 
from 338 K to 284 K in their calculation. Their methods also involve 
calculation of $\mu^*$ that is not treated in the Eliashberg form as well
as other methodological differences, and these differences make direct 
comparison with other reports difficult. Still, the relative effects of energy 
variation of N(E) are clear.

This DOS variation issue extends to the calculation of phonon frequencies. 
The phonon self-energy involves electron-phonon scattering in which a phonon 
is absorbed, scattering an electron from E$_k <$ E$_F$ to E$_k + \omega_q >$ E$_F$. 
Most methods of calculating phonon frequencies do not include effects of the
density of available initial or final states in this energy region being variable. Thus calculation
of phonon spectra will need to be re-evaluated for situations such as that 
imposed by H$_3$S, and the associated non-adiabatic corrections considered.

\section{Scenarios for room temperature}
The foregoing section indicates that 200 K superconductivity has been achieved 
with an effective DOS around ${\cal N}(E_F) \approx$ 0.44/eV-f.u. compared to a peak 
value between the two vHs around 0.7/eV-f.u. This means as mentioned above 
that the theory needs refining, as already noted by Flores-Livas {\it et al.}, 
to determine just how much is understood quantitatively and what features 
may require more attention.

This difference in N(E) versus ${\cal N}$(E) has larger and much more
positive implications. 
The fact that a reduced effective value of ${\cal N}(E_F)$ should be used
for H$_3$S also provides the glass-half-full viewpoint: a much larger value of 
${\cal N}(E_F)$ and therefore T$_c$ may be achievable in this or similar 
systems. Suppose that the two vHs can be moved apart, each by (say) 0.5 eV, 
leaving a value N(E$_F)$ $\approx$0.7/eV-f.u. between. Then the Lorentzian 
smearing will have much less effect. 
There is also the question of increasing the {\it magnitude} of N(E) at the peak, {\it i.e.}
N(E$_{vhs})$. This value depends on the effective 
masses at the vHs, but also on the volume of the region in which the quadratic
effective mass representation holds. If it holds in an ellipsoidal region 
defined by 
\begin{eqnarray}
\sum_j \frac{\hbar^2 k_j^2}{2m_j} < G_c^2,
\end{eqnarray}
the DOS from this region is
\begin{eqnarray}
N(E) \propto m_{th}^{3/2} [\alpha G_c - \beta \frac{|E-E_{vhs}|}{G_c} + ...],
\end{eqnarray}
where $\alpha$ and $\beta$ are numbers of order unity.
The value N(E$_{vhs})$ at the vHs is proportional to the thermal mass and to 
the radius $G_c$ of the
region of quadratic dispersion, and the decrease away from the vHs (the second term) is
inversely proportional to $G_c$. There will be additional smooth contributions from
outside this region, of course. However, increasing $m_{th}$ and the region
of quadratic dispersion is favorable for increasing ${\cal N}(E)$ in the vHs
region, and hence increasing T$_c$.
These observations seems to implicate the {\it topology} of the Fermi surfaces, 
rather than more conventional electronic structure characteristics such as
relative site energies and hybridization strengths. 

Numerical examples are illuminating. Suppose that the DOS peak can be widened
so that ${\cal N}(E_F) \approx$ N(E$_F$) = 0.70/ev-f.u. as outlined just above,
rather than the reduced  effective value of 0.45eV/f.u. that gives, experimentally,
T$_c$=200K.
With $\Omega =$ 1300 K and $\mu^*$=0.15 as above, 
$\lambda_{exp}$=2.17 is required to account for T$_c$ = 200K. For a
0.70/0.44 = 1.55 larger value of N(E$_F$), $\lambda$=3.38 and we find T$_c$=277K --
room temperature in a cool room. The increase in effective ${\cal N}(E_F)$ we have
assumed is ambitious but not outlandish, given the calculated spectrum
of $Im{\bar 3}m$ H$_3$S. It is clearly worthwhile to explore other H-rich
compounds for higher critical temperatures.

Of course, the increase in ${\cal N}(E_F)$ will give 
additional renormalization (softening) of the phonons. However, the modes 
are very stiff even with $\lambda$=2.1, so this may not be a 
major effect.  Note that decreasing $\Omega$ increases $\lambda$ but
decreases the energy scale prefactor in T$_c$, one reason why increasing
$\lambda$ by decreasing frequencies is rarely a profitable means of increasing
T$_c$. If $\Omega$, the prefactor in the Allen-Dynes T$_c$ equation, is softened 
by 10\% without change in matrix elements, $\lambda$ increases by 20\%
while T$_c$ increases by only 4\%.
Evidently softening of hard phonons is a
minor issue when looking for higher T$_c$ in this range of $\lambda$. 
This behavior was formalized by Allen and Dynes,\cite{alldyn} who obtained
the rigorous strong coupling limit
\begin{eqnarray}
T_c \rightarrow 0.18\sqrt{\lambda<\omega^2>} = 0.18\sqrt{N(E_F)<I^2>/M},
\end{eqnarray}
where $\omega$ is expressed in kelvins. The last expression is strictly
true only for an elemental superconductor, think of the electron-ion
matrix element $I$ and mass $M$ as those of H for H$_3$S. 

\section{Summary}
In this paper we first  addressed the electronic structure and
especially the delicate van Hove singularity induced spectrum, bonding
characteristics and effect of S, and the charge density of states near
the Fermi level, more directly than has been done before. The occurrence 
of two closely spaced van Hove singularities is definitely a central
issue for the properties of H$_3$S. We list some of the main points.

$\bullet$ At the most basic level, why is H$_3$S superconducting at 200
K? It is because both $\lambda$ is large but, more importantly, 
the characteristic phonon frequency
$\Omega$ is very high. This reminds one of the Allen-Dynes limit for
strong coupling, 
\begin{equation}T_c \rightarrow 0.18 \sqrt{\lambda \Omega^2} 
                    \rightarrow 0.18 \sqrt{N(E_F)<I^2>/M}.
\end{equation} 
Though not yet in this limit, this provides
the right picture -- one can check that keeping all fixed except for
$\Omega$ and then varying it, the change in T$_c$ is minor because the change
in prefactor $T_c\propto \Omega$ is compensated by $\lambda \propto
\Omega^{-2}.$   

$\bullet$ 
Sulfur $3p$ states hybridizing with hydrogen $1s$ is crucial in producing the
strong large scale structure
in N(E) within $\pm$5 eV of the Fermi level, and in leaving
E$_F$ at the top of a peculiarly sharp peak between two vHs. The van Hove points on the 
constant energy surfaces that define the peak in N(E) were identified,
finding they lie on opposite ends of a line of Fermi surface ``ripping apart''
with energy varying between the two van Hove singularities. This region
of very low velocity electrons affects a significant fraction of the zone.
It is unclear how replacing S with other elements will affect the
electronic structure near E$_F$, but small changes may have large
effects. Ge {\it et al.} have noted that alloying 7-10\% of P with S
moves $E_F$ to the peak in N(E), within the virtual crystal approximation
which does not account for alloy disorder broadening. In any case,
strong coupling smearing as discussed here will nullify this apparent gain.

$\bullet$ The fine structure  and energy variation of $N(E)$ near the Fermi
level must be taken into account to obtain quantitative results
for $\alpha^2F$, $\lambda$, and T$_c$. The energy dependence of N(E) may even affect
calculation of phonon frequencies, though this is untested so far.

$\bullet$ The closely spaced van Hove singularities 
very near the Fermi level have been shown to have significance, both on 
the detailed theory of H$_3$S but, as importantly, on the question of 
whether T$_c$ can be increased in related materials.
Sulfur and the specific $Im{\bar 3}m$ structure are important for high $T_c$
though other elements will need to be studied to learn more about  precisely why. 

$\bullet$ The prospect for
increased T$_c$  is affirmative -- it will require only evolutionary changes 
of the electronic structure to achieve room temperature superconductivity, 
though the road to this goal requires study, and additional insight into 
the origins of van Hove singularities may be important. Increasing the
vHs effective masses, or increasing  the volume within which quadratic
dispersion holds, will increase N(E) at the vHs energy. Structural
or chemical changes that affect the electronic structure rather modestly
may lead to significant increase in the effective (broadened) density
of states at E$_F$. Other studies
have suggested that substitution of some sulfur with chemically related
elements may increase T$_c$. Altogether, the prospects of achieving
increased critical temperatures are encouraging. 

$\bullet$ An issue that is almost untouched is a deeper understanding,
or rather an understanding at all, of electron-ion
matrix elements $<I_H^2>$ -- what contributes to strong electron-H atom scattering, 
and what degrades this scattering. These matrix elements  are the same 
that determine resistivity in the normal state; notably most of the 
best superconductors have high resistivities. Further study should
address the EPI matrix elements.

\section{Acknowledgments}We acknowledge helpful discussions with, and
comments on the manuscript from, A. S. Botana, B. M. Klein, M. J. Mehl, 
and D. A. Papaconstantopoulos. This work was supported by National
Science Foundation award DMR-1207622.

\appendix
\section{Details of van Hove singularities}
In FPLO, which we used initially to identify the effective masses at each vHs, 
the vHs energies lie at -0.43 eV and -0.11 eV, {\it i.e.} the Fermi level 
lies 110 meV above the upper vHs. In Wien2k, the vHs lie at -0.20 eV and 
+0.05 eV; E$_F$ lies just below the upper vHs, thus within the peak. 
The values of N(E$_{vhs})$ at the upper vHs are 0.555/eV-f.u. (FPLO)
and 0.630/eV-f.u. (Wien2k), reflecting the difference in thermal masses.
Pseudopotential results may differ by somewhat more than do these two methods,
which are usually in excellent agreement. 
\begin{figure}[!]
\includegraphics[width=0.8\columnwidth]{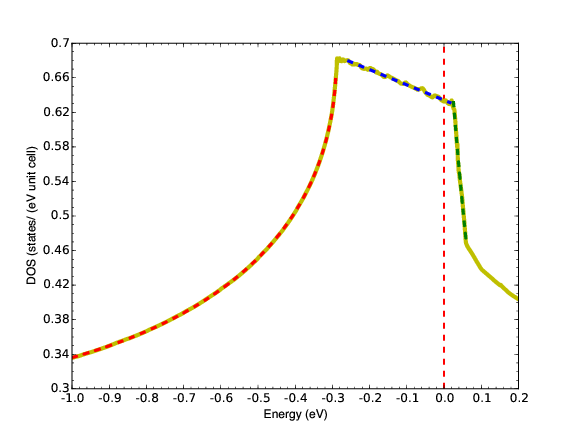}
\vskip 2mm
\includegraphics[width=0.8\columnwidth]{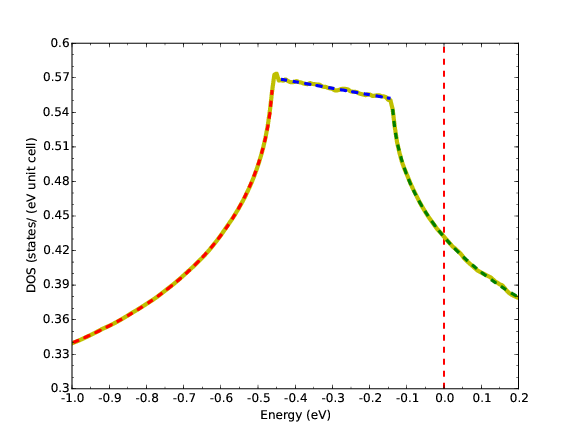}
\caption{Expanded view of the van Hove singularity region for WIEN2k (top panel) 
and FPLO (bottom panel) calculations.  In this energy range, there are two 
prominent van-Hove-singularities within 0.25 to 0.30 eV of one another, with
the top one lying very near E$_F$. The red dashed lines mark the Fermi level. 
Red, blue and green dashed lines are the fitted DOS using Eq. ~\ref{eqn1}. 
$N(E_f)$ from FPLO is 0.44 states/eV, while from WIEN2k it is 0.64 states/eV.}
\end{figure}

\section{Minimal tight-binding model}
A minimal tight-binding model for $H_3S$ has been constructed using Slater-Koster two-center 
hopping parameters. To simplify notation, we denote the sulfur $3s$ orbital as $S$, $3p$ 
orbitals as $P$, and hydrogen $1s$ orbital as $s$. The subscripts indicate the neighbor of 
the second site relative to the first. The Slater-Koster parameters, provided in  
Table I, are discussed in the text. 

\begin{figure}[!ht]
\centering
\includegraphics[width=0.3\columnwidth]{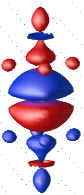}
\includegraphics[width=0.3\columnwidth]{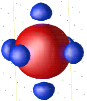}
\includegraphics[width=0.3\columnwidth]{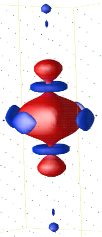}
\caption{Left: Sulfur $3p_z$ Wannier function, showing strong mixing
with H $1s$ states above and below. Center: Sulfur $3s$ Wannier 
function, with minor hybridization to neighboring $1s$ orbitals. 
Right: Hydrogen centered Wannier function, revealing strong hybridization
of H $1s$ with S $3p$ above and below,, as well as mixing with 
neighboring H $1s$ orbitals along the other two axes.}
\label{fig:my_label}
\end{figure}

Selected S-K matrix elements are provided that, by permutation, will allow 
construction of the tight binding model in the form we have used. For the 
row and column indices, 1 is S S,
2-4 are S P, and 5-7 are the H s orbitals.
\begin{eqnarray}
    H_{11} & = & (SS\sigma)_0+2(SS\sigma)_2(\cos k_x+\cos k_y+\cos k_z)\\ \nonumber
    H_{12} & = & 2i(SP\sigma)_2\sin k_x\\ \nonumber
    H_{15} & = & 2(Ss\sigma)_1\cos \frac{k_x}{2}\\ \nonumber
H_{22} & = & 2(\widetilde{PP})_1[\cos \frac{k_x+k_y+k_z}{2}\\ \nonumber
       &  & +\cos \frac{k_x+k_y-k_z}{2}+\cos \frac{k_x-k_y+k_z}{2} \\ \nonumber
       &  & +\cos \frac{k_x-k_y-k_z}{2}] + 2 (PP\sigma)_2 \cos k_x\\ \nonumber
H_{26 }& = &-(sP\sigma)_2\sqrt{2} i (\sin \frac{k_x-k_z}{2}+\sin\frac{k_x+k_z}{2})
                 \\ \nonumber
H_{55} & = & (ss\sigma)_0+2(ss\sigma)_4\cos k_x + \\ \nonumber
       & & 2 (ss\sigma)_4^\prime \cos k_y + 2(ss\sigma_4)^\prime\cos k_z
                      \\ \nonumber
H_{56} & = & 2(ss\sigma)_1 \cos \frac{k_z}{2}\\ \nonumber
\end{eqnarray}
Sulfur $P_x$ to $P_y$ hopping is small to the nearest neighbor and vanishes
by symmetry for the second neighbor, so $H_{23}$=0 at this level of the model. 
The dispersion obtained from the model Hamiltonian gives a reasonable,
but not quantitatively accurate, representation of the DFT bands.

\section{H$_3$H electronic structure}
The density of states and band structure of the model compound H$_3$H discussed 
in the text are presented here. This ``compound'' is actually simple cubic hydrogen, 
but presented in the H$_3$S type cell to facilitate comparison. The occupied
bandwidth is 15 eV. In bcc and fcc structures,\cite{papa2} there is no vHs below 
$E_F$. The conclusion is that sulfur has a momentous impact on the electronic 
structure within 5 eV of the Fermi level.
\begin{figure}[!th]
\centering
\includegraphics[width=0.8\columnwidth]{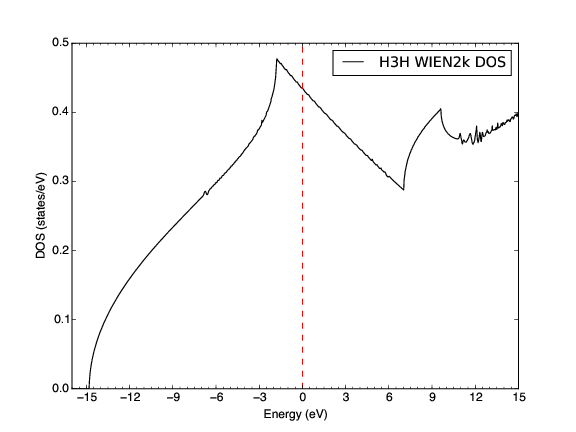}
\caption{H$_3$H density of states using WIEN2k with R$_H$ = 1.39$a_{\circ}$. R$_H$K$_{max}$ is set to 6 to be consistent with the H$_3$S calculations. The k-mesh was 100x100x100 with 22750 k-points in the irreducible wedge, and the tetrahedron method was used in generating the density of states. }
\label{H3Hdos}
\end{figure}

\begin{figure}
\centering
\includegraphics[width=0.7\columnwidth]{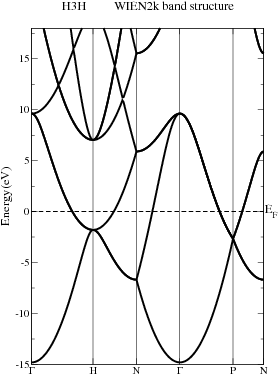}
\caption{H$_3$H band structure from WIEN2k, using
the cell of H$_3$S and calculational parameters as in the caption to Fig. \ref{H3Hdos}.}
\label{H3Hbands}
\end{figure}

\section{Spectral redistribution including $M_k(\omega)$}
Here the lower frequency region of the smearing of the spectral density is outlined.
We consider the regime where  $dM_k(\omega)/d\omega = -\lambda_k$, which holds
up to $\sim\Omega$ though perhaps not as far as $\Gamma = (\pi/2)\lambda\Omega$,
before the renormalization (mass renormalization) begins to ``burn off.''
Then returning to Eq. (4) before neglecting $M_k$,
\begin{eqnarray}
A(\omega)&=&\sum_k \frac{\Gamma_k(\omega)/\pi}
            {[\omega-(E_k-\mu)-M_k(\omega)]^2
            + \Gamma_k(\omega)^2}   \\ \nonumber
        &=&  \sum_k \frac{\Gamma_k(\omega)/\pi}
            {[(1+\lambda_k)\omega-(E_k-\mu)]^2
            + \Gamma_k(\omega)^2}   \\ \nonumber
        &\approx& \sum_k \frac{\Gamma_k(\omega)/\pi(1+\lambda)}
            {(\omega-\xi_k)]^2
            + \Gamma_k(\omega)^2/(1+\lambda)}   \\ \nonumber
       &\rightarrow& (1+\lambda)^{-3/2} \int d\xi N(\xi)  \frac{\tilde{\Gamma}/\pi}
            {(\omega-\xi)^2
            + \tilde{\Gamma}^2}.
\end{eqnarray}
The identity 1 = $\int \delta(\xi- \xi_k) d\xi$ has been inserted, $\xi_k = 
(E_k - \mu)/(1+\lambda_k)$, and in the
last line $k$-dependence other that through $\xi_k$ has been averaged.

In this expression the width  $\tilde{\Gamma} = \Gamma/\sqrt{1+\lambda}$ has been decreased
by mass renormalization by a factor of 1.8 for $\lambda$=2-2.2, reducing smearing. However, 
the overall magnitude has been decreased by $(1+\lambda)^{-3/2} \sim$ 1/6, a very
large expulsion of spectral weight from low energy by the mass enhancement. This
expression provides
additional insight into the substantial redistribution of low energy spectral weight
beyond the Lorentzian broadening (in the text) that takes over for $\omega > \Omega$.
Calculations of the self-energy for
sulfur hydrides\cite{Durajski} show that $M(\omega)$
differs only moderately from its low frequency form for $\omega$  
up to roughly the 2$\Omega$ range.

\end{document}